\documentclass[a4paper,onecolumn,11pt,accepted=2022-02-07]{quantumarticle}
\pdfoutput=1
\usepackage[utf8]{inputenc}
\usepackage[english]{babel}
\usepackage[T1]{fontenc}
\usepackage{amsmath,amsthm,amssymb,dsfont}
\usepackage{hyperref}
\usepackage[nameinlink]{cleveref}

\newtheorem{theorem}{Theorem}[section]
\newtheorem{definition}[theorem]{Definition}
\newtheorem{fact}[theorem]{Fact}
\newtheorem{claim}[theorem]{Claim}
\newtheorem{corollary}[theorem]{Corollary}
\newtheorem{lemma}[theorem]{Lemma}

\usepackage{complexity}
\usepackage{braket}

\renewcommand{\set}[1]{\left\{ #1 \right\}}
\newcommand{\abs}[1]{\left| #1 \right|}

\newcommand{\npower}[1]{{#1}^{\otimes n}}
\newcommand{\norm}[1]{\left\Vert {#1} \right\Vert}
\renewcommand{\C}{\mathbb{C}}
\renewcommand{\E}{\mathbb{E}}
\newcommand{\F}{\mathbb{F}}

\newcommand{\ind}{\mathds{1}}

\newcommand{\THR}{\mathsf{THR}}
\newcommand{\THRpn}{\THR_{pn}}
\newcommand{\Maj}{\mathsf{Maj}}
\newcommand{\diam}{\mathsf{diam}}

\begin{document}

\title{Lower Bounds on Stabilizer Rank}

\author{Shir Peleg}
\affiliation{Blavatnik School of Computer Science, Tel Aviv University, Tel Aviv, Israel}
\email{shirpele@tauex.tau.ac.il}
\thanks{The research leading to these results has received funding from the  Israel Science Foundation (grant number 514/20) and from the Len Blavatnik and the Blavatnik Family foundation.}
\author{Amir Shpilka}
\email{shpilka@tauex.tau.ac.il}
\affiliation{Blavatnik School of Computer Science, Tel Aviv University, Tel Aviv, Israel}
\thanks{The research leading to these results has received funding from the  Israel Science Foundation (grant number 514/20) and from the Len Blavatnik and the Blavatnik Family foundation.}
\author{Ben Lee Volk}
\email{benleevolk@gmail.com}
\affiliation{Efi Arazi School of Computer Science, Reichman University, Herzliya, Israel}
\thanks{Part of this work was done while at the Department of Computer Science, University of Texas at Austin, USA, Supported by NSF Grant CCF-1705028.}

\maketitle

\begin{abstract}
The \emph{stabilizer rank} of a quantum state $\psi$ is the minimal $r$ such that $\left| \psi \right \rangle = \sum_{j=1}^r c_j \left|\varphi_j \right\rangle$ for $c_j \in \mathbb{C}$ and stabilizer states $\varphi_j$. The running time of several classical simulation methods for quantum circuits is determined by the stabilizer rank of the $n$-th tensor power of single-qubit magic states.

We prove a lower bound of $\Omega(n)$ on the stabilizer rank of such states, improving a previous lower bound of $\Omega(\sqrt{n})$ of Bravyi, Smith and Smolin \cite{BSS16}. Further, we prove that for a sufficiently small constant $\delta$, the stabilizer rank of any state which is $\delta$-close to those states is $\Omega(\sqrt{n}/\log n)$. This is the first non-trivial lower bound for approximate stabilizer rank.

Our techniques rely on the representation of stabilizer states as quadratic functions over affine subspaces of $\mathbb{F}_2^n$, and we use tools from analysis of boolean functions and complexity theory. The proof of the first result involves a careful analysis of directional derivatives of quadratic polynomials, whereas the proof of the second result uses Razborov-Smolensky low degree polynomial approximations and correlation bounds against the majority function. 
\end{abstract}

\section{Introduction}

The conventional wisdom is that quantum computers are more powerful than classical computers. Among other reasons, this belief is supported by the fact that quantum computers are able to efficiently solve problems such as integer factorization \cite{Shor97}, which are believed by some to be hard for classical computers; by provable black box separations \cite{Simon97, Grover96, BV97, RT19}; and by quantum computers' advantage in solving certain sampling problems that are deemed intractable for classical computers under well established complexity theoretic conjectures \cite{AA13}.

There is, however, very little that we can \emph{unconditionally} prove with regard to the impossibility of efficiently simulating quantum computers using classical computers. Indeed, barring a computational complexity theoretic breakthrough, such as --- at the very least --- separating $\P$ from $\PSPACE$, we can't hope to prove general and unconditional impossibility results.

Nevertheless, it remains an interesting and important problem to prove lower bounds on the running time of certain restricted types of simulation techniques for quantum circuits. One such result is a lower bound of Huang, Newman and Szegedy \cite{HNS20}, who prove unconditional exponential lower bounds for a subclass of simulators they call \emph{monotone simulators}, which includes many, but not all, of the known simulation techniques.

Simulation algorithms based on \emph{stabilizer rank decompositions} for quantum circuits dominated by Clifford gates \cite{BSS16, BG16, BBCCGH19} is a recent powerful class of algorithms for classically simulating quantum circuits (which is not covered by the lower bound of \cite{HNS20}). The computational cost of these algorithms is dominated by a certain natural algebraic and complexity-theoretic rank measure for quantum states, which we now define.

\subsection{Clifford Circuits, Magic States and Stabilizer Rank}
\label{sec:clifford}

Clifford circuits are quantum circuits which only apply Clifford gates (for background on the Clifford group and the definitions of the type of gates considered in this paper, see \Cref{app:notation}). Equivalently, such circuits only use CNOT, Hadamard, and phase gates. This is an important class of quantum circuits which, by the Gottesman-Knill theorem \cite{Gottesman97, AG04}, \emph{can} be efficiently simulated (on, say, the input $\ket{0^n}$) by a classical algorithm. This highly non-obvious theorem follows from the fact that such circuits can only maintain certain states known as \emph{stabilizer states}. These can be succinctly represented, and it is easy to track the state and update the succinct representation after any application of a Clifford gate.

Adding $T$ gates (we refer again to \Cref{app:notation} for the definition) on top of the Clifford gates results in a universal quantum gate set, that is, a set which can approximate every unitary operation. It is then possible, using a simple gadget-based transformation, to ``push the $T$ gates to the inputs'' and obtain an equivalent circuit, of roughly the same size, which only uses Clifford operations, and is given, as additional auxiliary inputs, sufficiently many copies of qubits in a so-called \emph{magic state} \cite{BK05, BSS16}. This transformation only increases the circuit size by a polynomial factor. For classical circuit complexity theorists, a useful albeit imperfect analogy is the fact that any size-$s$ circuit can be simulated by a \emph{monotone} circuit of size polynomial in $s$, which is given as additional inputs the $n$ negations of the input variables $x_1, \ldots, x_n$.

Two examples for such magic states are $\ket{H} = \cos(\pi/ 8 ) \ket{0} + \sin(\pi/8) \ket{1}$ and $\ket{R} = \cos(\beta)\ket{0} + e^{i \pi /4} \sin(\beta)\ket{1}$, where $\beta = \frac{1}{2}\arccos(1/\sqrt{3})$ \cite{BK05}.\footnote{The state $\ket{R}$ is often called $\ket{T}$. However, to avoid confusion, we follow the notation of \cite{BSS16}, and reserve the notation $\ket{T}$ for a different state. For a handy reference to our notation, see \Cref{app:notation}.} This suggests the possibility of simulating a general quantum circuit by decomposing $\ket{\npower{H}}$ or $\ket{\npower{R}}$ as a linear combination of stabilizer states.

More formally, $\ket{\varphi}$ is a \emph{stabilizer state} if $\ket{\varphi} = U\ket{0^n}$ where $U$ is an $n$-bit Clifford unitary (see also Equation~\eqref{eq:stabilizer-state} and the following paragraph).  The \emph{stabilizer rank} of a state $\ket{\psi}$, denoted $\chi(\psi)$, is the minimal integer $r$ such that
\[
\ket{\psi} = \sum_{j=1}^r c_j \ket{\varphi_j},
\]
where for every $1 \le j \le r$, $\ket{\varphi_j}$ is a stabilizer state and $c_j \in \C$.

For any $n$-qubit state, the stabilizer rank is at most $2^n$. Interestingly, much smaller upper bounds can be shown for the the stabilizer rank of $\ket{\npower{H}}$: Bravyi, Smith and Smolin \cite{BSS16} proved that $\chi(H^{\otimes 6}) \le 7$ which implies that $\chi(\npower{H}) \le 7^{n/6} \le 2^{0.468 n}$. Bravyi, Smith and Smolin \cite{BSS16} then use this identity to obtain simulation algorithms for circuits with a small number of $T$ gates, whose running time is much faster than the trivial brute-force simulation. A slightly faster algorithm was presented by Kocia who proved that $\chi(H^{\otimes 12}) \le 47$ \cite{Kocia20}, and upper bound was further improved by Qassim, Pashayan and Gosset \cite{QPG21} who proved that $\chi(\npower{H}) = O(2^{\alpha n})$ for $\alpha =\frac{1}{4} \log_2 3 \le 0.3963$.

When simulating quantum circuits, it is often enough, for all intents and purposes, to obtain an approximation of their output state. Thus, it's natural to define a similar approximation notion for stabilizer rank. The $\delta$-approximate stabilizer rank of $\ket{\psi}$, denoted $\chi_{\delta}(\psi)$, is defined as the minimum of $\chi(\varphi)$ over all states $\ket{\varphi}$ such that $\norm{\psi - \varphi }_2 \le \delta$ \cite{BBCCGH19}. By considering approximate stabilizer decomposition of $\ket{H^{\otimes n}}$, improved simulation algorithms were obtained by Bravyi and Gosset \cite{BG16}.

A natural question is then what is the limit of such simulation methods. As the running time of the simulation scales with the stabilizer rank, an upper bound which is polynomial (in $n$) on $\chi(\npower{H})$ or $\chi(\npower{R})$ will imply that $\BPP=\BQP$ and even (by simulating quantum circuits with postselection) $\P=\NP$ \cite{BBCCGH19}, and thus seems highly improbable.\footnote{This implication holds up to uniformity issues having to do with finding the decomposition of $\ket{H^{\otimes n}}$ as a linear combination of stabilizer states. However, these complexity classes collapses are not believed to hold even in the non-uniform world, and further, by the Karp-Lipton theorem, a non-uniform collapse also implies a collapse of the polynomial hierarchy in the uniform world.} Much stronger hardness assumptions than $\P \neq \NP$, such as the exponential time hypothesis, imply that $\chi(\npower{H}) = 2^{\Omega(n)}$ \cite{MT19, HNS20}.

However, the starting point of this discussion was our desire to obtain \emph{unconditional} impossibility results, and thus we are interested in provable lower bounds on $\chi(\npower{H})$ and $\chi_\delta (H^{\otimes n})$, and similarly for $\npower{R}$.

While it's easy to see, using counting arguments, that the stabilizer rank of a random quantum state would be exponential, it is a challenging open problem to prove super-polynomial lower bounds on the rank of $\ket{\npower{H}}$ or for other explicit states.
Bravyi, Smith and Smolin proved that $\chi(\npower{H}) = \Omega(\sqrt{n})$. In this paper, we improve this lower bound, and also prove the first non-trivial lower bounds for approximate stabilizer rank.

\subsection{Our results: Improved Lower Bounds on Stabilizer Rank and Approximate Stabilizer Rank}
\label{sec:results}

Our first result is an improved lower bound on $\chi(\npower{H})$ and $\chi(\npower{R})$.

\begin{theorem}
\label{thm:intro:exact}
$\chi(\npower{H}) = \Omega(n)$, and similarly, $\chi(\npower{R}) = \Omega(n)$.
\end{theorem}

As we remark in \Cref{sec:technique}, proving super-linear lower bound on $\chi(\npower{H})$ will solve a notable open problem in complexity theory. We discuss this challenge, as well us some barriers preventing our technique from proving super-linear lower bounds, in \Cref{sec:open}.

The result of \Cref{thm:intro:exact} can be immediately adapted to prove the same lower bounds on the $\delta$-approximate stabilizer rank for exponentially small $\delta$. We are, however, interested in much coarser approximations, and we are able to prove a meaningful result even for $\delta$ being a small enough positive constant.

\begin{theorem}
\label{thm:intro:approximate}
There exists an absolute constant $\delta>0$ such that $\chi_{\delta}(\npower{H}) = \Omega(\sqrt{n}/\log n)$, and similarly $\chi_{\delta}(\npower{R}) = \Omega(\sqrt{n}/\log n)$.
\end{theorem}

By definition, the stabilizer rank of any two states which are Clifford-equivalent is the same, and thus the lower bounds of \Cref{thm:intro:exact} and \Cref{thm:intro:approximate}, while stated as lower bounds on the ranks of $\ket{\npower{H}}$ and $\ket{\npower{R}}$ hold for any state which is Clifford-equivalent to them, even up to a phase.

\subsection{Technique: Stabilizer States as Quadratic Polynomials}
\label{sec:technique}

The original proof of the Gottesman-Knill Theorem used the stabilizer formalism and tracked the current state of the circuit by storing the generators of the subgroup of the Pauli group which stabilizes the state, and updating them after each application of a Clifford operation. It turns out, however, that there is an alternative succinct representation of stabilizer states, using their amplitudes in the computational basis $\set{\ket{x}}_{x \in \F_2^n}$ \cite{DD03,VandenNest10}. This representation also leads to an alternative proof of the theorem, as explained in \cite{VandenNest10}.

If $\ket{\varphi}$ is a stabilizer state then (up to normalization)
\begin{equation}
\label{eq:stabilizer-state}
\ket{\varphi} = \sum_{x \in A} i^{\ell(x)} (-1)^{q(x)} \ket{x}
\end{equation}
where $A \subseteq \F_2^n$ is an affine subspace, $\ell(x)$ is an $\F_2$-linear function and $q(x)$ is a quadratic polynomial over $\F_2$. The amplitudes of $\ket{\npower{H}}$ and $\ket{\npower{R}}$ are also easy to compute. For example, recall that $\ket{H} = \cos(\pi/8)\ket{0} + \sin(\pi/8) \ket{1}$, and thus
\[
\ket{\npower{H}} = \sum_{x \in \F_2^n} \cos(\pi/8)^{n - |x|} \sin(\pi/8)^{|x|} \ket{x},
\]
where $|x|$ denotes the Hamming weight of $x$. 

It is convenient to recast this problem as a problem about functions on the boolean cube in the following natural way. For an $n$-qubit state $\ket{\psi}$ we associate a function $F_\psi : \F_2^n \to \C$ such that $F_{\psi}(x)$ equals the amplitude of $\ket{x}$ when writing $\ket{\psi}$ in the computational basis. In this formulation, our ``building blocks'' are \emph{stabilizer functions}, i.e., functions of the form
\[
\varphi(x) = i^{\ell(x)} (-1)^{q(x)} \ind_A
\]
where $A$ is an affine subspace, $\ind_A$ is the indicator function of $A$ (i.e., $\ind_A(x)=1$ if $x \in A$ and zero otherwise), $\ell$ is a linear function and $q$ is a quadratic polynomial. Let $H_n$ denote the function associated with $\ket{\npower{H}}$. We would like to show that in any decomposition
\[
H_n (x) = \sum_{j=1}^r c_j \varphi_j (x) = \sum_{j=1}^r c_j i^{\ell(x)} (-1)^{q_j(x)} \ind_{A_j} (x)
\]
where  $c_j \in \C$ and $\varphi_j(x)$ are stabilizer functions, $r$ must be large.

Our techniques for showing that use tools from the analysis of boolean functions and from complexity theory. In \Cref{sec:prev} we recall some similar questions that have arisen in complexity theory.

For the proof of \Cref{thm:intro:exact}, we show that if $f$ is a function of stabilizer rank at most, say, $n/100$, then it is possible to find two vectors $x,y \in \F_2^n$ such that the Hamming weight of $x$ is very small, the Hamming weight of $y$ is very large, and $f(x) = f(x+y)$. Since $|x+y| \ge |y|-|x|$, for the correctly chosen parameters we get that $|x+y| > |x|$, which leads to a contradiction if $f=H_n$, since $H_n$ takes different value on each layer of the Hamming cube.

To find such $x$ and $y$, given a decomposition $\sum_{j=1}^r c_j i^{\ell(x)} (-1)^{q_j} \ind_{A_j}$ with $r \le n/100$, we find $x,y$ such that $\ell_j(x) = \ell_j(x+y)$, $q_j(x) = q_j(x + y)$ and $\ind_{A_j}(x) = \ind_{A_j}(x+y)$ for all $j \in [r]$.

Observe that for a fixed $y\in \F_2^n$ and a quadratic polynomial $q(x)$, the equation $q(x) = q(x+y)$ is an affine linear equation in unknowns $x$. Thus, denoting $\Delta_y(q) = q(x) + q(x+y)$ (this is also called the directional derivative of $q$ with respect to $y$), we get a system of affine linear equations $\set{\Delta_y(q_j) = 0}_{j \in [r]}$ in $x$, which, assuming $r$ is small, has many solutions (assuming it is solvable at all).

The additional requirements $\ell_j(x) = \ell_j(x+y)$ and $\ind_{A_j}(x) = \ind_{A_j}(x+y)$ make things more complicated. However, using an averaging argument and by again utilizing the fact that $r$ is relatively small, we are able to find a large affine subspace $U$ of vectors which satisfy those equations, and then we analyze the above system of linear equations over the affine subspace $U$.

In order to satisfy the conditions on the Hamming weights of $x$ and $y$ we use Kleitman's theorem \cite{Kleitman66} which gives an upper bound on the size of sets of the boolean cube with small diameter, as well as some elementary linear algebra. The full proof of \Cref{thm:intro:exact} appears in \Cref{sec:exact}.

The proof of \Cref{thm:intro:approximate} follows a different strategy. Starting from a state $\ket{\psi}$ of rank $r$ which is $\delta$-close to $\ket{\npower{H}}$ for some small enough constant $\delta > 0$, we show how to use $\ket{\psi}$ in order to construct an $\F_2$-polynomial of degree $O(r \log r)$ which $(1-\varepsilon)$-approximates the majority function on $m=\Omega(n)$ bits. By a well known correlation bound of Razborov and Smolensky \cite{Razborov87, Smolensky87, Smolensky93}, this implies that $r = \Omega(\sqrt{n}/\log n)$.

We now explain how to obtain this polynomial approximating the majority function. Let $p = \sin^2(\pi/8) = 0.146\dots$. Instead of majority, it is convenient to first consider the function $\THRpn$ which is 1 on all inputs $x$ whose Hamming weight is at least $pn$, and zero otherwise. Note that this function is trivial to approximate under the uniform distribution by the constant 1 polynomial, but the approximation question becomes meaningful when considering $B(n,p)$, the binomial distribution with parameter $p$ on the $n$-dimensional cube. This is useful since the $L_2$ mass of the vector $\ket{\npower{H}}$ is distributed according to this distribution. In particular it is heavily concentrated on coordinates $x$ such that $|x| = pn \pm O(\sqrt{n})$, and a state $\ket{\psi}$ which is $\delta$-close to $\ket{H^{\otimes n}}$ must contain in almost all of these coordinates values which are very close to those of $\ket{\npower{H}}$. It is then possible to obtain from $\psi$ a boolean function $f$ which approximates the function $\THRpn$. We observe that a restriction $g$ of $f$ to a random set of $2pn$ coordinates will approximate the majority function, and further, assuming $\ket{\psi}$ has stabilizer rank $r$, and using standard techniques again borrowed from Razborov and Smolensky, $g$ itself can be approximated by a polynomial $\tilde{g}$ of degree $O(r \log r)$. It follows that $\tilde{g}$ approximates the majority function over $2pn$ bits. The full proof of \Cref{thm:intro:approximate} appears in \Cref{sec:approximate}.

\subsection{Related Work}
\label{sec:prev}

As mentioned above, the previous best lower bound was an $\Omega(\sqrt{n})$ lower bound for exact stabilizer rank of $\ket{\npower{H}}$ proved by Bravyi, Smith and Smolin \cite{BSS16}. Stronger lower bounds are known in restricted models. As mentioned by \cite{BSS16} (see also Lemma 2 in \cite{BG16}), for every stabilizer state $\ket{\varphi}$ it holds that $\abs{\braket{\varphi|\npower{H}}} \le 2^{-\Omega(n)}$ which immediately implies an exponential lower bound in the case that the coefficients $c_j$ are bounded in magnitude (in particular, this holds if the states in the decomposition are orthogonal). It is worth noting that by Cramer's rule, in any rank $r$ decomposition the coefficients $c_j$ can be taken to be of magnitude at most exponential in $n$ and $r$.

Bravyi et al.\ \cite{BBCCGH19} present a different restricted model in which they prove an exponential lower bound.

Related questions have been considered before in complexity theory. The so called ``quadratic uncertainty principle'' \cite{OpenProblemsSimons, Williams18} is a conjecture which states that in any decomposition of the AND function as a sum
\begin{equation}
\label{eq:quadratic}
\sum_{j=1}^r c_j (-1)^{q_j (x)},
\end{equation}
for quadratic functions $\set{q_j}_{j \in [r]}$ and $c_j \in \C$, $r = 2^{\Omega(n)}$. The best lower bound known is $r \ge n/2$ (see \cite{Williams18}). Note that since in the stabilizer rank case we allow functions of the form $(-1)^{q} \cdot \ind_A$ for affine subspaces $A$, the model we consider in this paper is stronger: in particular the AND function itself is a stabilizer function and its stabilizer rank is 1.

Williams \cite{Williams18} has constructed, for every positive integer $k$, a function $f_k \in \NP$ which requires $r = \Omega(n^k)$ in any decomposition as in \eqref{eq:quadratic}. It remains, however, an intriguing open problem to construct boolean function in $\P$ which requires a super-linear number of summands.

We remark that proving super linear lower bounds on the stabilizer rank of $\ket{\npower{H}}$ will solve this problem. Indeed, as mentioned above, the stabilizer rank model is even stronger, and thus lower bounds carry over to weaker models. Furthermore, even though $H_n$ itself is not a boolean function, $\ket{H}$ is Clifford-equivalent (up to an unimportant phase) to $\ket{T} := \frac{1}{\sqrt{2}} (\ket{0} + e^{i \pi /4 } \ket{1})$ (see \cite{BSS16}), which implies that the stabilizer rank of $\ket{\npower{H}}$ equals the stabilizer rank of $\ket{\npower{T}}$. Denoting $T_n$ the function associated with $\npower{T}$, it is now evident that $T_n(x)$ depends only on $|x| \bmod 8$, and therefore
\[
T_n = \sum_{j=0}^7 b_j M_j(x),
\]
where for $j \in \set{0,...,7}$, $b_j \in \C$ and $M_j : \F_2^n \to \set{0,1}$ is a boolean function such that $M_j(x) =1$ if and only if $|x|=j \bmod 8$. Thus, a super-linear lower bound on the stabilizer rank of $\ket{\npower{H}}$ will imply a super-linear lower bound on the rank of the (boolean) mod 8 function.

Following the initial publication of this work, our results were reproved using different techniques. Labib \cite{Labib21} used higher-order Fourier analysis in order to prove a result similar to \Cref{thm:intro:exact}, and extended it to qudits of any prime dimension. Lovitz and Steffan \cite{LS21} proved nearly identical lower bounds for exact and approximate stabilizer rank using number-theoretic techniques.

\subsection{Open Problems}
\label{sec:open}

While \Cref{thm:intro:exact} improves upon the previous best lower bound known, we are unfortunately unable to prove super-polynomial or even super-linear lower bounds on $\chi(\npower{H})$ or $\chi(\npower{R})$. Further, our techniques seem incapable of proving super-linear lower bounds, as they extend to any representation of $H_n$ as an arbitrary function of $r$ stabilizer functions, and not necessarily a linear combination of them.

As mentioned in \Cref{sec:prev}, it seems that a first step could be proving super-linear lower bounds for the quadratic uncertainty principle problem. A different approachable open problem is to improve our lower bound on the $\delta$-approximate stabilizer rank to be closer to $\Omega(n)$. This could perhaps be easier assuming $\delta$ is polynomially small in $n$.

\paragraph*{Acknowledgements} The third author would like to thank Andru Gheorghiu for introducing him to the notion of stabilizer rank.

\section{Preliminaries}
\label{sec:prelim}

\subsection{General Notation}
As mentioned in the introduction, it is often convenient to speak about functions on the boolean cube rather than quantum states. For an $n$-qubit state $\ket{\psi} = \sum_{x \in \F_2^n} c_x \ket{x}$, the associated function $F_{\psi} : \F_2^n \to \C$ is defined as $F_{\psi}(x) = c_x$.

The $L_2$ norm of the function $F : \F_2^n \to \C$ is then the same as the norm of the corresponding vector, i.e., $\norm{F} = \left( \sum_{x \in \F_2^n} |F(x)|^2 \right)^{1/2}$.

A function $\varphi : \F_2^n \to \C$ is called a \emph{stabilizer function} if there exists an $n$-variate linear function $\ell(x)$, an $n$-variate quadratic polynomial $q(x) \in \F_2[x_1, \ldots, x_n]$ and an affine subspace $A \subseteq \F_2^n$ such that $\varphi(x) = i^{\ell(x)} (-1)^{q(x)} \ind_A$, where $\ind_A$ denotes the characteristic function of $A$. As shown in \cite{DD03, VandenNest10}, stabilizer functions indeed correspond to stabilizer states up to normalization (which has no effect on the stabilizer rank).

The \emph{stabilizer rank} of a function $F : \F_2^n \to \C$, denoted $\chi(F)$, is the minimal $r$ such that there exist $c_1, \ldots, c_r \in \C$ and stabilizer functions $\varphi_1, \ldots, \varphi_r$ such that $F(x) = \sum_{j=1}^r c_j \varphi_j(x)$.

For a vector $x \in \F_2^n$ we denote by $|x|$ its Hamming weight. We denote by $\Maj_m : \F_2^m \to \F_2$ the $m$-bit majority function, that is $\Maj_m(x) = 1$ if and only if $|x| \ge m/2$.

\begin{definition}
Let $A \subset \F_2^n$. The \emph{diameter} of $A$, denoted $\diam(A)$, is defined as
\[
\max_{u,v \in A} d(u,v) = \max_{u,v \in A} |u+v|.
\]
Here $d(u,v)$ denotes the Hamming distance of $u$ and $v$.
\end{definition}

Kleitman \cite{Kleitman66} proved that sets of small diameter cannot be too large.

\begin{theorem}[\cite{Kleitman66}]\label{thm:klei}
Let $A\subset \F_2^n$ such that $\diam(A) \le 2k$ for $k < n/2$. Then,
\[
|A| \leq \sum_{j=0}^{k} \binom{n}{j} \leq 2^{H_2\left(\frac{k}{n} \right) n},
\]
where $H_2(p) = -p \log_2 p -(1-p) \log_2 (1-p)$ is the binary entropy function.
\end{theorem}
This result is obviously tight as shown by the example of the set of all vectors of Hamming weight at most $k$.


\subsection{Linear Algebraic Facts}

Recall that an affine subspace $U \in \F_2^n$ is a the set of solutions to a system of affine equations, i.e., a system of the form $Mx = b$ for some $M \in \F_2^{k \times n}$ and $b \in \F_2^k$. Every affine subspace can be written as $U = u + U_0$ for $u \in \F_2^n$ and a linear subspace $U_0 \subseteq \F_2^n$. In our terminology, linear subspaces are in particular affine subspaces (and similarly, linear functions are a special case of affine functions).

We record the following useful facts.

\begin{fact}\label{fa:affine-perp}
	Let $U\subsetneq \F_2^n$ be an affine subspace, and let $v \in \F_2^n \setminus U$. Then there is an affine function $a(x) : \F_2^n \to \F_2$ such that $a(v)=1$ and for every $u \in U$, $a(u)=0$. \end{fact}

\begin{fact}\label{fa:1-iso-affine}
Let $U_1, U_2 \subseteq \F_2^n$ be affine subspaces such that $U_1\cap U_2\neq \emptyset$. Then
\[
\dim(U_1 + U_2)= \dim(U_1)+\dim(U_2) - \dim(U_1 \cap U_2).
\]
\end{fact}

\begin{claim}\label{cl:fixed-sol}
	Let $U \subseteq \F_2^n$ be an affine subspace, with $\dim(U) = n-k>0$. There exists a subset $S\subset [n]$, of size $|S|=n-k$ such that for every $v \in \F_2^{n-k}$  there is $u \in U$ with $u|_S=v$ (where $u|_S$ denotes the restriction of $u$ to the coordinates indexed by $S$).
\end{claim}
\begin{proof}
$U$ is the set of solutions for an equation $Mx=b$ for a matrix $M \in \F_2^{k\times n}$ and $b\in \F_2^k$. The fact that $\dim(U) = n-k$ implies that $M$ has rank $k$, and there is a $k\times k$ non-singular submatrix $M'$ of $M$. Denote by $S$ the columns of $M$ that do not appear in $M'$. For every $v \in \F_2^{n-k}$, fixing $x|_{S} = v$ in the equation $Mx=b$ gives a system of equations $M'x'=b'$ in the set of remaining $k$ unknowns $x'$, which has a solution since $M'$ is non-singular.
\end{proof}

\begin{corollary}\label{cl:small-wt}
Let $U \subseteq \F_2^n$ be an affine subspace with $\dim(U) = n-k>0$. Then, there exists $u \in U$ with $|u| \leq k$.
\end{corollary}  
\begin{proof}
Follows immediately from applying \Cref{cl:fixed-sol} with $v =0$. 
\end{proof}

Finally, we define the directional derivative of a quadratic function over $\F_2$.

\begin{definition}\label{def:derivative}
Let $q \in \F_2[x_1, \ldots, x_n]$ be a polynomial of degree $2$. Let $0 \neq y \in \F_2^n$. The \emph{directional derivative} of $q$ in direction $y$ is defined to be the function
\[
\Delta_y(q)(x) := q(x) + q(x+y) \in \F_2[x_1, \ldots, x_n].
\]
Observe that for every $y$, $\Delta_y(q)$ is an affine function in $x$.
\end{definition}

\section{A Lower Bound for Exact Stabilizer Rank}
\label{sec:exact}

In this section we prove \Cref{thm:intro:exact}. We first present the main lemma of this section.

\begin{lemma}
\label{lem:exact-weight}
Let $F : \F_2^n\to \C$ be a function of stabilizer rank $r$ such that $r \le n/100$. Then, there exist $y, z \in \F_2^n$ such that $|y| \neq |z|$ and $F(y) = F(z)$.
\end{lemma}

\Cref{thm:intro:exact}, which we now restate, is an immediate corollary of \Cref{lem:exact-weight}.

\begin{theorem}
\label{thm:exact}
Let $\ket{B}$ be either $\ket{H}$ or $\ket{R}$. Then $\chi (\npower{B}) = \Omega(n)$.
\end{theorem}

\begin{proof}
In the case where $\ket{B}=\ket{H}$, the associated function $F_H : \F_2^n \to \C$ is defined by $F_H(x) = \cos(\pi/8)^{n-|x|}\sin(\pi/8)^{|x|}$. If $\ket{B}=\ket{R}$,
the associated function $F_R : \F_2^n \to \C$ is defined by $F_R(x) =  \cos(\beta)^{n-|x|}(e^{i \pi /4} \sin(\beta))^{|x|}$ where $\beta = \arccos(1/\sqrt{3})/2$.

It is immediate to verify that for every $y,z \in \F_2^n$ of different Hamming weight those functions attain different values. Thus, by \Cref{lem:exact-weight}, their stabilizer rank is at least $n/100$.
\end{proof}

We turn to the proof of \Cref{lem:exact-weight}.

\begin{proof}[Proof of \Cref{lem:exact-weight}]
Let $F : \F_2^n \to \C$ be a function of stabilizer rank at most $r \le n/100$, i.e.,
\[
F(x) = \sum_{j=1}^r c_j i^{\ell_j(x)} (-1)^{q_j(x)} \ind_{A_j} (x),
\]
where for every $j \in [r]$, $\ell_j$ is a linear function, $q_j$ is a quadratic function, and $A_j \subseteq \F_2^n$ is an affine subspace. 

To prove the statement of the lemma, we will show that there exist $y,z \in \F_2^n$ such that $|y| < |z|$ and for every $j \in [r]$ all of the following hold:
\begin{enumerate}
\item $\ell_j(y) = \ell_j(z)$
\item $\ind_{A_j}(y) = \ind_{A_j}(z)$
\item $q_j(y) = q_j(z)$.
\end{enumerate}

The first two items are handled by the following claim, which shows that there is a large affine subspace satisfying both conditions.

\begin{claim}
\label{cl:linear-and-affine}
There's an affine subspace $U \subseteq \F_2^n$ of dimension at least $n-3r$ such that for every $j \in [r]$ and for every $u_1, u_2 \in U$, $\ell_j(u_1) = \ell_j(u_2)$
and $\ind_{A_j}(u_1) = \ind_{A_j}(u_2)$.
\end{claim}

We defer the proof of \Cref{cl:linear-and-affine} to the end of this proof. Write $U = u + U_0$ where $u \in \F_2^n$ and $U_0 \subseteq \F_2^n$ is a \emph{linear} subspace. The next claim handles the third item above.

\begin{claim}\label{cl:heavy-solution}
There exists $v \in U_0$ with $|v| \ge 2n/3$ such that the system of equations \[ \set{q_j(x) = q_j(x+v)}_{j \in [r]} \] (in unknowns $x$) has a solution in $U$.
\end{claim}
 
We postpone the proof of this claim as well, and now explain how it implies the result. Let $v \in U_0$ as promised in \Cref{cl:heavy-solution}. The set of solutions in $U$ to the system of affine equations
\begin{equation}
\label{eq:quadratic-deriv-system}
\set{q_j(x) = q_j(x+v)}_{j \in [r]} = \set{\Delta_v(q)(x) = 0}_{j \in [r]}
\end{equation}
is non-empty (by \Cref{cl:heavy-solution}), and thus by \Cref{fa:1-iso-affine}, the set of solutions in $U$ to \eqref{eq:quadratic-deriv-system} is an affine subspace $V \subseteq U$ of dimension at least $n-4r$.

By \Cref{cl:small-wt}, there is $y \in V$ with $|y| \leq 4r$. Set $z = y + v$, so that $q_j(y) = q_j(y + v) = q_j(z)$ for all $j \in [r]$. Observe that $z \in U$, since \Cref{cl:heavy-solution} promises that $v \in U_0$. Thus $y$ and $z$ attain the same values on $\ell_j$ and $\ind_{A_j}$ for all $j \in [r]$ as well. Finally note that $|y| \le 4r$ whereas
\[
|z| = |y + v| \ge |v| - |y| \ge \frac{2n}{3} - 4r > 4r. \qedhere
\]
\end{proof}
It remains to prove \Cref{cl:linear-and-affine} and \Cref{cl:heavy-solution}.

\begin{proof}[Proof of \Cref{cl:linear-and-affine}]
Let $V_1 \subset \F_2^n$ be the linear subspace defined by the system of equations $\set{\ell_j = 0}$ for all $j \in [r]$. It holds that $\dim(V_1) \geq n-r>0$.

Consider now the map $E : V_1 \to \set{0,1}^r$, defined by 
\[
E(x) = (\ind_{A_1}(x),\ldots, \ind_{A_r}(x)).
\]
By the pigeonhole principle, there is $\alpha \in \set{0,1}^r$ with $|E^{-1}(\alpha)|\geq 2^{\dim{V_1}-r}\geq 2^{n-2r}$. Let $S$ be the support of $\alpha$, that is, the set of indices $j \in [r]$ such that $\alpha_j=1$. We have that
\[
E^{-1}(\alpha) = \left( \left( \bigcap_{j\in S} A_j \right) \setminus \left( \bigcup_{j \not\in S} A_j \right) \right) \cap V_1 \subseteq \left(\bigcap_{j\in S} A_j \right) \cap V_1
\]
(for notational convenience, if $S=\emptyset$, then $\bigcap_{j\in S} A_j = \F_2^n$).

Let $V_2 =  \left(\bigcap_{j\in S} A_j \right) \cap V_1$. Then $V_2$ is an affine subspace, and $|V_2| \ge |E^{-1}(\alpha)| \ge  2^{n-2r}$, so $\dim(V_2) \ge n-2r > 0$.

Pick now an arbitrary $x_0 \in E^{-1} (\alpha)$. Thus, $x_0 \in V_2$, and for every $j \not\in S$, $x_0 \not\in A_j$. By \Cref{fa:affine-perp}, for every $j \not\in S$ there is an affine equation $a_j$ such that $a_j(x_0)=1$ and for all $x \in A_j$, $a_j(x)=0$. Let
\[
U = \set{x \in V_2 : \text{for all } j \not\in S, a_j(x)=1}.
\]
Then $U$ is an affine subspace (as it is defined by at most $r$ additional affine constraints on $V_2$), and it is non-empty (since $x_0 \in U$). By \Cref{fa:1-iso-affine}, it follows that $\dim(U) \ge n-2r-r = n-3r$. Further, for every $x \in U$ and $j \in [r]$, it holds that $\ell_j(x) = 0$ and
\[
\ind_{A_j}(x) = \begin{cases}
1 & j \in S \\
0 & j \not\in S
\end{cases}
\]
which completes the proof.
\end{proof}

We finish the section by proving \Cref{cl:heavy-solution}.

\begin{proof}[Proof of \Cref{cl:heavy-solution}]
Consider the map $\Gamma : U \to \set{0,1}^{r}$ defined by
\[
\Gamma(x) = (q_1(x), \ldots, q_r(x)).
\]
For every $\alpha \in \set{0,1}^r$, let $\Gamma_\alpha = \set{x_1 + x_2: x_1, x_2 \in \Gamma^{-1}(\alpha)}$. Observe that for every $\alpha$, $\Gamma_\alpha \subseteq U_0$. Furthermore, for every $v \in \Gamma_\alpha$, the set of affine equations \[
\set{\Delta_v(q_j)(x) = 0}_{j \in [r]},
\]
in unknowns $x$, has a solution in $U$. Indeed, $v = x_1 + x_2$ where $x_1, x_2 \in \Gamma^{-1}(\alpha)$, and thus $q_j(x_1) = q_j(x_2) = q_j(x_1+v)$ for every $j \in [r]$, which implies that $x_1$ is a solution. 

In order to finish the proof we need to show that there is $\alpha \in \set{0,1}^r$ and $v \in \Gamma_{\alpha}$ such that $|v| \geq \frac{2n}{3}$. By the pigeonhole principle there is $\alpha_0 \in \set{0,1}^r$ such that $|\Gamma^{-1}(\alpha_0)| \ge |U|/2^r = 2^{n-4r}$. Observe that the maximal Hamming weight of an element in $\Gamma_{\alpha_0}$ equals the diameter of the set $\Gamma^{-1}(\alpha_0)$.

By \Cref{thm:klei} (for $k=n/3$), the size of every set of diameter $2n/3$ is at most $2^{H_2(1/3) n} \leq 2^{0.92n}$. Since $r \le n/100$, $|\Gamma^{-1}(\alpha_0)| > 2^{0.95n}$, so $\diam(\Gamma^{-1}(\alpha_0)) \ge 2n/3$, and there is $v \in \Gamma_{\alpha_0}$ of weight at least $2n/3$.
\end{proof}

\section{A Lower Bound for Approximate Stabilizer Rank}
\label{sec:approximate}

In this section we prove \Cref{thm:intro:approximate}. In \Cref{sec:threshold-to-majority}, we show how to obtain, given a function $f$ that approximates the function $\THRpn$ (with respect to the binomial distribution on the $n$-dimensional cube with parameter $p$,  $B(n,p)$), a random restriction of $f$ which approximates the majority function over $m= 2\lfloor pn \rfloor$, bits with respect to the uniform distribution.\footnote{In what follows, in order to help with the readability of the argument we often omit the floor and ceiling signs. For example, we'll use ``$pn$'' to refer to $\lfloor pn \rfloor$. We reintroduce floor and ceiling signs in cases where there is a chance of confusion.}
 In  \Cref{sec:low-rank-to-threshold}, we construct, given a state $\ket{\psi}$ that is $\delta$ close to either $\ket{\npower{H}}$ or $\ket{\npower{R}}$, a boolean function $f_\psi$ that approximates $\THRpn$. In \Cref{sec:rs-approximation} we then show how to get low-degree polynomial approximations to restrictions of $f_\psi$, which, as we specify in \Cref{sec:lb-corr}, completes the proof.

\subsection{A Reduction from Threshold Functions to Majority}
\label{sec:threshold-to-majority}

Let $0<p<1/2$. Recall that $\THRpn(x)$ equals $1$ if $|x| \ge pn$ and $0$ otherwise. In this section we prove that given any function $f: \F_2^n \to \F_2$ that approximates  $\THRpn$ with respect to $B(n,p)$, we can find a function $g$, which is a restriction of $f$ to $2pn$ random coordinates, which approximates the majority function on those bits with respect to the uniform distribution.

In anticipation of the next section, when considering approximations for $\THRpn$ we will work with a slightly different notion of approximation than approximation with respect to $B(n,p)$, which we now explain.

Let $L_k = \set{x \in \F_2^n \mid |x|=k}$ denote the $k$-th layer of the boolean cube.  We say that a function $f: \F_2^n \to \F_2$ is \emph{$\varepsilon$-wrong} on $L_k$ (with respect to $\THRpn$) if the fraction of elements $x \in L_k$ such that $f(x) \neq \THRpn(x)$ is at least $\varepsilon$.

We say that $f$ \emph{$(\varepsilon, \gamma)$-approximates $\THRpn$} if $f$ is $\varepsilon$-wrong on at most a $\gamma$ fraction of the layers $L_k$ for $k \in [pn - \lceil 5\sqrt{2pn} \rceil, pn + \lceil 5\sqrt{2pn} \rceil]$.

For the rest of the proof we will always set $\varepsilon = \gamma = 0.01$.

Since $B(n,p)$ is heavily concentrated on layers $L_k$ with $k \in [pn - O(\sqrt{n}), pn + O(\sqrt{n})]$, and for every $k$ in that range, $\Pr_{x \sim B(n,p)} [x \in L_k] = \Theta(1/\sqrt{n})$, this notion and the notion of approximation with respect to $B(n,p)$ are in fact very similar, up to the precise choice of constants.

\begin{lemma}
\label{lem:restriction-to-maj}
Let $0<p < 1/2$ be an absolute constant, and let $f : \F_2^n \to \F_2$ be a boolean function that $(0.01, 0.01)$-approximates $\THRpn$. For every $D \subseteq [n]$ of size $m:=2pn$, let $g_D : \F_2^m \to \F_2$ be the function obtained from $f$ by fixing all input bits outside of $D$ to $0$. Then there exists $D_0$ such that for $g:=g_{D_0}$, $\Pr_{x \in \F_2^m} [g(x) = \Maj_m(x)] \ge 3/4$, where $x$ is chosen according to the uniform distribution.
\end{lemma}

\begin{proof}
Let $m = 2pn$. For every $D \subseteq [n]$ of size $m$, let $g_D$ be the function obtained from $f$ by fixing all input bits outside of $D$ to $0$. It will be convenient to consider $g_D$ as a function whose domain is $\F_2^m$ using some bijection between $D$ and $[m]$. Every $x \in \F_2^n$ which is zero on coordinates outside of $D$ then corresponds to a unique $\bar{x} \in \F_2^m$, and vice versa.

We will now pick $D$ uniformly at random among all subsets of $[n]$ of size $m$, so that $g_D$ is a random restriction of $f$.

We say $x \in \F_2^n$ \emph{survives} $D$ if the set of indices $j \in [n]$ such that $x_j = 1$ is contained in $D$. The probability that $x \in L_k$ survives $D$ is $\binom{m}{k}/\binom{n}{k}$.

For an input $x \in  \F_2^n$, we say $x$ is \emph{correct} if $f(x) = \THRpn(x)$, and incorrect otherwise. If $x$ is correct and survives, then $ \Maj_m(\bar{x}) = \THRpn(x) = f(x) = g_D(\bar{x})$.

Let $X_k$ be a random variable, which denotes the number of incorrect inputs $x \in L_k$ that survive $D$, and
\[
X = \sum_{k=pn - \lceil 5 \sqrt{2pn} \rceil}^{pn + \lceil 5 \sqrt{2pn}\rceil } X_k.
\]
By the assumption, for at least $0.99$ fraction of the layers $L_k$, the number of incorrect $x$'s is at most $0.01 \binom{n}{k}$, and thus for each such layer $L_k$ for $k \in [pn - \lceil 5\sqrt{2pn} \rceil, pn + \lceil 5\sqrt{2pn}\rceil]$, $\E_D[X_k] \le 0.01 \binom{m}{k}$.  We call such layers \emph{good}. For the rest of the layers, which we call \emph{bad}, obviously $\E_D[X_k] \le \binom{m}{k}$. The total number of  layers in the interval $[pn - \lceil 5\sqrt{2pn} \rceil, pn + \lceil 5\sqrt{2pn}\rceil]$ is at most
\[
2 \cdot (5\sqrt{2pn} +1 ) +1 \le 11\sqrt{2pn},
\]
and thus the number of bad layers
 is at most $0.01 \cdot 11\cdot \sqrt{2pn}$. Further, for every $k$, $\binom{m}{k} \le \frac{1}{\sqrt{m}} \cdot 2^m$.

Therefore,
\begin{align*}
\E_D[X] & = \sum_{L_k\text{ is good}} \E_D[X_k] + \sum_{L_k\text{ is bad}} \E_D[X_k] \\
& \le \sum_{L_k\text{ is good}} 0.01 \binom{m}{k} + 0.01 \cdot 11\sqrt{m} \cdot \frac{1}{\sqrt{m}} \cdot 2^m \\
& \le 0.01 \cdot \left( \sum_k \binom{m}{k} \right) + \frac{11}{100} \cdot 2^m \le \frac{12}{100} \cdot 2^m.
\end{align*}

In particular, there is some $D_0$ such that the number if incorrect $x$'s in layers $[pn - \lceil 5\sqrt{2pn} \rceil, pn+ \lceil 5\sqrt{2pn} \rceil ]$ that survive $D_0$ is at most $\frac{12}{100} 2^m$. Let $g := g_{D_0}$. We now claim that $g$ and $\Maj_m$ agree on more than $3/4$ of the inputs in $\F_2^m$. 

First, By the Chernoff bound, the number of vectors $\bar{x} \in \F_2^m$ whose Hamming weight is \emph{not} in the range
\[
[pn - \rceil 5\sqrt{2pn} \lceil, pn + \lceil 5\sqrt{2pn} \rceil] = [m/2 - \lceil 5\sqrt{m} \rceil, m/2+\lceil 5\sqrt{m} \rceil],
\]
is at most $2e^{-((5/\sqrt{pn})^2\cdot pn /6)}\cdot 2^m\leq \frac{1}{15}2^m$. On these inputs we have no guarantee. By the choice of $D_0$, the number of $\bar{x}$'s such that $|\bar{x}| \in [m/2 - \lceil 5\sqrt{m} \rceil, m/2+\lceil 5\sqrt{m} \rceil]$ and $g(x) \neq \Maj(x)$ is at most $\frac{12}{100} 2^m$. It follows that $g(\bar{x}) \neq \Maj_m(\bar{x})$ on less than $\frac{1}{4} \cdot 2^m$ inputs.
\end{proof}

\subsection{From Stabilizer Decompositions to Threshold Functions}
\label{sec:low-rank-to-threshold}

Let $\ket{B} = \alpha \ket{0} + \beta \ket{1}$ with $|\alpha|^2 + |\beta|^2=1$. Let $p = |\beta|^2$ and suppose that $0<p<1/2$. Let $F_B : \F_2^n \to \C$ be the function associated with $\ket{\npower{B}}$, i.e., $F_B (x) = \alpha^{n-|x|} \beta^{|x|}$. 

In this section we prove that if $\psi : \F_2^n \to \C$ is such that $\chi(\psi) \le r$ and $\norm{\psi - F_B} \le \delta$, then it is possible to construct a boolean function $f_\psi$ that $(0.01,0.01)$-approximates $\THRpn$. In \Cref{sec:rs-approximation}, we will prove that if $\chi(\psi) \le r$, $f_\psi$ has low degree polynomial approximations.

From here on, $\delta$ will denote a sufficiently small constant, which may depend on $\ket{B}$ and its parameters (i.e., $\delta$ is some function of $p$), but does \emph{not} depend on $n$. Since we are  interested in the case $\ket{B}=\ket{H}$ or $\ket{B}=\ket{R}$, $\delta$ can be taken to be some small universal constant.

For $k \in [n]$, let $m_k := |\alpha^{n-k} \beta^{k}|$ denote the absolute value of $F_B$ on the $k$-th layer.
Let $w_k = m_k^2 =  p^k(1-p)^{n-k}$ and $W_k = \binom{n}{k} w_k$ the total mass on the $k$-th layer, with respect to $B(n,p)$. Let $\eta =  \frac{|\beta|}{|\alpha|}$. Observe that by assumption, $0 < \eta < 1$.

Suppose $\psi : \F_2^n \to \C$ is such that $\chi(\psi) \le r$ and $\norm{\psi - F_B} \le \delta$. 
We define a boolean function $f_\psi: \F_2^n \to \F_2$ as follows:\footnote{Observe that if $|x|<|x'|$ then $|\psi(x)|>|\psi(x')|$.} 
\begin{equation}
\label{eq:boolean-def}
f_\psi (x) = \begin{cases}
1 & \text{if }  | \psi(x) | \le \left( \frac{1+\eta}{2} \right) m_{pn-1}  \\
0 & \text{otherwise}
\end{cases}
\end{equation}
The intuition for the definition is that, since $\norm{\psi-F_B} \le \delta$, we expect $\psi(x)$ to be very close to $F_B(x)$ for most inputs $x$. For every such $x$, $f_{\psi}$ will correctly compute $\THRpn$. Further, inputs $x$ such that $f_{\psi} (x) \neq \THRpn(x)$ correspond to inputs $x$ such that $|\psi(x)-F_B(x)|$ is large. Assuming there are many such $x$'s will lead to a contradiction to the assumption that $\norm{\psi-F_B} \le \delta$.

\begin{lemma}
\label{lem:thrpn-approx}
Let $\psi : \F_2^n \to \C$ be a function such that $\norm{\psi - F_B} \le \delta$ for a sufficiently small $\delta$. Let $f_\psi$ the boolean function defined as in \eqref{eq:boolean-def}. Then $f_\psi$ $(0.01, 0.01)$-approximates $\THRpn$.
\end{lemma}

We begin with the following calculation.

\begin{claim}
\label{cl:difference-on-wrong-inputs}
Suppose $x \in \F_2^n$ is such that $f_\psi(x) \neq \THRpn(x)$ and $|x|=k$. Then $|\psi(x) - F_B(x)|^2 \ge w_k \cdot \left( \frac{1-\eta}{2} \right)^2$.
\end{claim}

\begin{proof}
Since $|x|=k$, $|F_B(x)|=m_k$.
Suppose first that $k \le pn-1$ so that $\THRpn(x)=0$. By assumption, $f_\psi(x)=1$, which implies that
\[
|\psi(x)| \le \left( \frac{1+\eta}{2} \right) m_{pn-1}.
\]
Observe that $m_k = (\eta^{-1})^{pn-1-k}m_{pn-1} \ge m_{pn-1}$ for $k \le pn-1$, and therefore by the triangle inequality
\begin{align*}
|\psi(x) - F_B(x)| \ge |F_B(x)| - |\psi(x)| & \ge m_{k} - \left( \frac{1+\eta}{2} \right) m_{pn-1} \\
& \ge m_k - \left( \frac{1+\eta}{2} \right) m_k = \left( \frac{1-\eta}{2} \right) m_{k},
\end{align*}
which implies the statement of the claim (for $k \le pn-1$) by squaring both sides.

If $k \ge pn$, then $\THRpn(x)=1$ which implies $f_\psi(x)=0$, i.e.,
\[
|\psi(x)| \ge \left( \frac{1+\eta}{2} \right) m_{pn-1}.
\]
Note that $m_k = \eta^{k-pn+1}m_{pn-1}$ and in particular $m_k \le \eta m_{pn-1}$ for all $k \ge pn$. Thus,
\begin{align*}
|\psi(x) - F_B(x)| \ge |\psi(x)| - |F_B(x)|  & \ge \left( \frac{1+\eta}{2} \right) m_{pn-1} - m_{k}  \\
& \ge \left( \frac{1+\eta}{2} \right) m_{pn-1} -  \eta m_{pn-1} = \left( \frac{1-\eta}{2} \right) m_{pn-1} \\
& \ge  \left( \frac{1-\eta}{2} \right) m_{k},
\end{align*}
which proves the lemma for this case as well.
\end{proof}

We use the following standard estimates on the concentration of the binomial distribution. Recall that $W_k = \binom{n}{k} p^k (1-p)^{n-k}$.

\begin{claim}
\label{cl:central-binomial-coeffs}
Let $C \in \mathbb{R}$. Then $W_{pn +C \sqrt{n}} = \Omega(1/\sqrt{n})$, where the constant hidden under the $\Omega$ notation depends on $C$ and $p$, but not on $n$.
\end{claim}

Observe that $C$ in the above claim may be negative. The proof is a direct application of Stirling's approximation. For completeness, we provide a crude estimate which suffices for us in \Cref{app:binomial}.

We are now ready to prove the main lemma of the section.

\begin{proof}[Proof of \Cref{lem:thrpn-approx}]
Let $k$ be a layer such that $f_\psi$ is $0.01$-wrong on $L_k$. By \Cref{cl:difference-on-wrong-inputs},
\[
\sum_{x \in L_k} |\psi(x) - F_B(x)|^2 \ge 0.01 \cdot \binom{n}{k} \cdot \left(\frac{1-\eta}{2}\right)^2 w_k = 0.01 \left(\frac{1-\eta}{2}\right)^2 W_k.
\]
Suppose, towards a contradiction, $f_\psi$ is $0.01$-wrong on more than $0.01$ fraction of the layers $k \in [pn - \lceil 5\sqrt{2pn} \rceil, pn + \lceil 5 \sqrt{2pn} \rceil]$, i.e., on more than $0.1\sqrt{2pn}$ layers. By \Cref{cl:central-binomial-coeffs}, for every such $k$,
$W_k \ge c/\sqrt{n}$ for some constant $c$ which does not depend on $n$. It follows that
\[
\norm{\psi - F_B} \ge 0.1 \sqrt{2pn} \cdot 0.01 \left(\frac{1-\eta}{2}\right)^2 \cdot \frac{c}{\sqrt{n}},
\]
which is a contradiction for $\delta < 0.001 \sqrt{2p} c  \left(\frac{1-\eta}{2}\right)^2$.
\end{proof}

\subsection{A Low Degree Polynomial Approximation}
\label{sec:rs-approximation}

In this section we show that for the function $f_\psi$ defined as in \eqref{eq:boolean-def}, and for any restriction $g_D$ of $f_\psi$ as in  \Cref{lem:restriction-to-maj}, the function $g_D$ has a polynomial approximating it, whose degree is at most $O(r \log r)$. To prove this we apply standard approximation techniques used for proving lower bounds for bounded depth circuits with modular gates, although in our case the details are somewhat simpler.

We begin with the following lemma that shows how to approximate indicator functions of affine subspaces with low degree polynomials.

\begin{claim}[\cite{Razborov87, Smolensky93}]
\label{cl:rs-approx-subspace}
Let $A \subseteq \F_2^m$ be an affine subspace. For every $t \in \mathbb{N}$, there exists a polynomial $P \in \F_2[x_1, \ldots, x_m]$ of degree at most $t$ such that $\Pr_{x \in \F_2^m} [P(x) \neq \ind_A(x)] \le 2^{-t}$.
\end{claim}

\begin{proof}
Since $A$ is an affine subspace, there exist $k \le m$ affine functions $a_1, \ldots, a_k$ such that $x \in A$ if and only if $a_j(x) = 0$ for every $j \in [k]$, or equivalently, $\ind_A(x) = \prod_{j=1}^k (a_j(x) + 1)$.

Let $D$ be a uniformly random subset of $[k]$ and $a_D = \sum_{j \in D} a _j $. Observe that for $x \in A$, $a_D(x) = 0$ with probability 1, whereas for $x \not \in A$, there is some $j \in [k]$ such that $a_j(x) = 1$ and hence $\Pr_D [ a_D (x) = 0] = 1/2$ (as $j$ is included in $D$ with probability $1/2$).

Hence, for $t \in \mathbb{N}$, define $P_{\mathbf{D}}(x) = \prod_{k=1}^t (a_{D_k} (x) + 1)$, where $\mathbf{D} = (D_1, \ldots, D_t)$ are chosen uniformly and independently. Then, $P_{\mathbf{D}}$ is a degree $t$ (random) polynomial, $P_{\mathbf{D}}(x) = 1$ for all $x \in A$, and for $x \not\in A$, $\Pr_{\mathbf{D}} [P_{\mathbf{D}}(x)=1] \le 2^{-t}$. In particular, in expectation $P_{\mathbf{D}}$ and $\ind_A$ disagree on at most $2^{m-t}$ of the inputs, which implies that there exists a choice of $\mathbf{D}'=(D_1, ..., D_t)$ such that $P:=P_{\mathbf{D}'}$ satisfies the properties required in the lemma.
\end{proof}
We now show how to approximate restrictions of the boolean function $f_\psi$.

\begin{lemma}
\label{lem:rs-approximation-function}
Let $F_B$ and $\psi$ be functions as in \Cref{sec:low-rank-to-threshold} and let $f_\psi$ defined as in \eqref{eq:boolean-def}. Let $D \subseteq [n]$ and denote $g:=g_D$ the restriction of $f_\psi$ obtained by setting variables outside of $D$ to $0$, as in \Cref{sec:threshold-to-majority}. Then, there is a polynomial $\tilde{g}$ of degree $O(r \log r)$ such that $\Pr_{\bar{x} \in \F_2^m} [g(\bar{x}) \neq \tilde{g}(\bar{x})] \le \frac{1}{20}$.
\end{lemma}

\begin{proof}
	Write
	\begin{equation}\label{eq:psi}
		\psi(x) = \sum_{j=1}^r c_j \varphi_j(x) = \sum_{j=1}^r c_j i^{\ell_j(x)} (-1)^{q_j(x)} \ind_{A_j}(x),
	\end{equation}
	where for every $j \in [r]$, $\ell_j$ is a linear function, $q_j$ a quadratic function, and $A_j$ an affine subspace.
	
For every $j \in [r]$, let $A'_j, \ell'_j, q'_j$ denote the projection of $A_j, \ell_j, q_j$ respectively, obtained by setting the coordinates outside of $D$ to zero. Observe that $A'_j \subseteq \F_2^m$ is an affine subspace, $\ell'_j$ an $m$-variate linear function over $\F_2$, and $q'_j$ an $m$-variate quadratic function over $\F_2$, and that
\[
g(\bar{x}) = \begin{cases}
1 & \text{if }  | \sum_{j=1}^r c_j \cdot i^{\ell'_j (\bar{x})} \cdot (-1)^{q'_j(\bar{x})} \cdot \ind_{A'_j}(\bar{x}) | \le \left( \frac{1+\eta}{2} \right) m_{pn-1} \\
0 & \text{otherwise.}
\end{cases}
\]
Let $h: \F_2^{3r} \to \F_2$ denote the following function:
\[
h(y_1, \ldots, y_r, z_1, \ldots, z_r, v_1, \ldots, v_r ) = \begin{cases}
1 & \text{if }  | \sum_{j=1}^r c_j \cdot i^{y_j} \cdot (-1)^{z_j} \cdot v_j | \le \left( \frac{1+\eta}{2} \right) m_{pn-1} \\
0 & \text{otherwise.}
\end{cases}
\]
(note that here $v_j \in \set{0,1}$ is considered as a real number).
Then \[
g(\bar{x}) = h(\ell'_1(\bar{x}), \ldots, \ell'_r(\bar{x}), q'_1(\bar{x}), \ldots, q'_r(\bar{x}), \ind_{A'_1}(\bar{x}), \ldots, \ind_{A'_r}(\bar{x})).
\]
For every $j \in [r]$, let $P_j$ be a polynomial of degree $O(\log(r))$ such that $\Pr_{\bar{x} \in \F_2^m} [P_j(\bar{x}) \neq \ind_{A'_j} (\bar{x})] \le \frac{1}{20r}$, as guaranteed by \Cref{cl:rs-approx-subspace}. Note that $h$ is a function on $3r$ boolean variables, and hence can be represented exactly by a polynomial of degree at most $3r$. As the $\ell'_j$'s have degree $1$ and $q'_j$'s degree $2$, it follows that
\[
\tilde{g}(\bar{x}) = h(\ell'_1(\bar{x}), \ldots, \ell'_r(\bar{x}), q'_1(\bar{x}), \ldots, q'_r(\bar{x}), P_1(\bar{x}), \ldots, P_r(\bar{x}))
\]
is a polynomial of degree $O(r \log r)$, and by the union bound
\[
\Pr_{\bar{x} \in \F_2^m} \left[ \tilde{g}(\bar{x}) \neq g(\bar{x}) \right] \le \Pr_{\bar{x} \in \F_2^m} \left[ \exists j \in [r] \text{ such that } P_j(x) \neq \ind_{A'_j(x)} \right] \le \frac{1}{20}. \qedhere
\]
\end{proof}

\subsection{A Lower Bound for Approximate Stabilizer Rank via Correlation Bounds}
\label{sec:lb-corr}

We now observe that the results of \Cref{sec:threshold-to-majority}, \Cref{sec:low-rank-to-threshold} and \Cref{sec:rs-approximation} imply our lower bounds. The final ingredient we require is the following correlation lower bound.

\begin{lemma}[\cite{Razborov87, Smolensky93}]
\label{lem:rs-deg-lb}
Let $f : \F_2^m \to \F_2$ be a boolean function such that 
\[
\Pr_{x \in \F_2^m} [f(x) = \Maj_m(x)] \ge \frac{2}{3}.
\]
Then $\deg(f) = \Omega(\sqrt{m})$.
\end{lemma}

We recall \Cref{thm:intro:approximate}
\begin{theorem}[Restatement of  \Cref{thm:intro:approximate}]
\label{thm:approximate}
Let $\ket{B}$ be either $\ket{H}$ or $\ket{R}$. Then, for a sufficiently small constant $\delta$, it holds that $\chi_\delta (\npower{B}) = \Omega(\sqrt{n}/\log n)$.
\end{theorem}

\begin{proof}
Let $\psi$ be a state such that $\norm{\psi - \npower{B}} \le \delta$. By \Cref{lem:thrpn-approx}, this implies that the boolean function $f := f_\psi$, as defined in \eqref{eq:boolean-def}, $(0.01, 0.01)$-approximates $\THRpn$. By \Cref{lem:restriction-to-maj} this implies that there exists a restriction of $f$, $g$, such that
\[
\Pr_{\bar{x}\in\F_2^m} [g(\bar{x}) \neq \Maj_m(\bar{x})] \le \frac{1}{4}
\]
for $m=2pn$. Further, by \Cref{lem:rs-approximation-function}, there is a polynomial $\tilde{g}$, of degree $O(r \log r)$, such that
\[
\Pr_{\bar{x}\in\F_2^m} [g(\bar{x}) \neq \tilde{g}(\bar{x})] \le \frac{1}{20}.
\]
It follows that
\[
\Pr_{\bar{x}\in\F_2^m} [\tilde{g}(\bar{x}) \neq \Maj_m(\bar{x})] \le \frac{1}{3}
\]
and thus, by \Cref{lem:rs-deg-lb}, $r \log r = \Omega(\sqrt{2pn})$, as the theorem states.
\end{proof}

\bibliographystyle{quantumurl}
\bibliography{quantum-ref}

\appendix

\section{The Clifford Group and Magic States}
\label{app:notation}

The purpose of this section is to provide a brief introduction to the Clifford group for readers who are unfamiliar with it. We shall not cover the entire background, motivation and various applications of this group in quantum computing and quantum information, but rather only provide the bare minimum of definitions needed to understand this work and its motivation. The book \cite{NC16} is good extensive reference on these topics, and in particular Sections 10.5.1 and 10.5.2 which deal with the stabilizer formalism. We also provide a notational reference to the various gates and magic states we consider in this paper.

\subsection{Pauli and Clifford Group}
\label{app:pauli}
The \emph{Pauli matrices} are three $2 \times 2$ complex unitary matrices defined as follows: 
\[
X = \begin{bmatrix}0 & 1 \\ 1 & 0\end{bmatrix}, \quad
Y = \begin{bmatrix}0 & -i \\ i & 0\end{bmatrix}, \quad
Z = \begin{bmatrix}1 & 0 \\ 0 & -1\end{bmatrix}.
\]
These matrices generate a subgroup of $2x2$ matrices of order 16, denoted by $P_1$ and called the single qubit Pauli group, that contains the elements 
\[
\set{\pm I, \pm i I, \pm X, \pm iX, \pm Y, \pm iY, \pm Z, \pm iZ}.
\]
The $n$-qubit Pauli group, denoted $P_n$ is defined as
\[
P_n = \set{ \sigma_1 \otimes \sigma _2 \otimes \cdots \otimes \sigma_n : \text{for all } j \in [n], \sigma_j \in P_1}.
\]

The Clifford group $\mathcal{C}_n$ can now be defined as the normalizer of $P_n$ in the group $U(n)$ of $n$-qubit unitary matrices. It is convenient, however, to consider $\mathcal{C}_n$ as a finite group, which is why it is usually defined modulo $U(1)$, i.e., we identify two matrices $U$ and $V$ if $U=c V$ for some $c \in \C$ with $|c|=1$ ($c$ is called a \emph{global phase}):
\[
\mathcal{C}_n := \set{U \in U(n) : U P_n U^{\dagger} = P_n} / U(1).
\]

It turns out that $\mathcal{C}_n$ has a set of generators which is very easy to describe. Every $U \in \mathcal{C}_n$ can be generated using the following simple set of gates:
\[
\text{CNOT} = \begin{bmatrix} 1 & 0 & 0 & 0 \\ 0 & 1 & 0 & 0 \\ 0 & 0 & 0 & 1 \\ 0 & 0 & 1 & 0 \end{bmatrix}, \quad
H = \frac{1}{\sqrt{2}} \begin{bmatrix}1 & 1 \\ 1 & -1\end{bmatrix}, \quad
S = \begin{bmatrix}1 & 0 \\ 0 & i\end{bmatrix}.
\]
$H$ is called the Hadamard gate and $S$ is called the phase gate. The set of \emph{stabilizer states} is the set of states $\varphi$ such that $\ket{\varphi} = U\ket{0^n}$.

Evidently, $\set{\text{CNOT},H,S}$ is thus not a universal quantum gate set. However, the set $\set{\text{CNOT},H,S, T}$, where
\[
T = \begin{bmatrix}1 & 0 \\ 0 & e^{i \pi /4} \end{bmatrix}
\]
is the so-called $\pi/8$ gate, \emph{is} universal.

\subsection{Magic States}
\label{app:magic}

As explained in \Cref{sec:clifford}, any circuit over the (universal) gate set  $\set{\text{CNOT},H,S, T}$ can be converted to a circuit of roughly the same size with only Clifford gates, which is given as additional inputs an ample supply of qubits in a \emph{magic state}. The two types of magic states defined by Bravyi and Kitaev \cite{BK05} are
\[
\ket{H} = \cos(\pi/8) \ket{0} + \sin(\pi/8) \ket{1}, \quad \text{and} \ket{R} = \cos(\beta) \ket{0} + e^{i \pi / 4} \sin(\beta) \ket{1},
\]
where $\beta = \arccos(1/\sqrt{3})/2$.

We say two $n$-qubit states $\psi$ and $\varphi$ are \emph{Clifford-equivalent} if $\ket{\psi} = U\ket{\varphi}$ for $U \in \mathcal{C}_n$. Up to a phase, state $\ket{H}$ is Clifford-equivalent to the state $\ket{T} = \frac{1}{\sqrt{2}} (\ket{0} + e^{i \pi / 4} \ket{1})$ (see \cite{BSS16}), and thus Clifford circuits provided with $\ket{\npower{H}}$ as auxiliary inputs have the same computational power as Clifford circuits provided with $\ket{\npower{T}}$.

\section{Proof of \texorpdfstring{\Cref{cl:central-binomial-coeffs}}{Claim 4.4}}
\label{app:binomial}

\begin{proof}[Proof of \Cref{cl:central-binomial-coeffs}]
Recall that by Stirling's approximation, $m! \sim \sqrt{2\pi m} \left( \frac{m}{e} \right)^m$. In particular, for large enough $n$,
\begin{align*}
 \binom{n}{pn}  &= \frac{n!}{(pn)! ((1-p)n)!}  \\
&\ge  \frac{1}{2} \frac{\sqrt{2 \pi n} \cdot (n/e)^n}{\sqrt{2 \pi (pn)} (pn/e)^{pn} \cdot \sqrt{2 \pi (1-p)n} \cdot ((1-p)n/e)^{(1-p)n}}.
\end{align*}
Thus,
\[
W_{pn} = \binom{n}{pn}p^{pn}(1-p)^{(1-p)n}  = \Omega(1/\sqrt{n}),
\]
 where the constant hidden under the $\Omega$ notation depends on  $p$.
Now, for $C>0$, we will show that $W_{pn} / W_{pn + C\sqrt{n}}= O(1)$ (where again, the constant depends on $C$ and $p$).
\begin{align*}
\frac{W_{pn}}{W_{pn + C\sqrt{n}}} & = \frac{\binom{n}{pn} p^{pn} (1-p)^{(1-p)n}}{\binom{n}{pn + C\sqrt{n}} p^{pn+C\sqrt{n}} (1-p)^{(1-p)n - C\sqrt{n}}} \\
& = \frac{(pn + C\sqrt{n}) \cdots (pn+1)}{((1-p)n) \cdots ((1-p)n - C\sqrt{n}+1)} \cdot \left( \frac{1-p}{p} \right)^{C\sqrt{n}} \\
& \le \left(  \frac{pn+C\sqrt{n}}{(1-p)n - C\sqrt{n}} \right)^{C\sqrt{n}} \cdot \left( \frac{1-p}{p} \right)^{C\sqrt{n}} \\
& = \frac{pn}{(1-p)n}^{C \sqrt{n}} \cdot \frac{\left( 1+\frac{C}{p\sqrt{n}}\right)^{C\sqrt{n}}}{\left( 1-\frac{C}{(1-p)\sqrt{n}}\right)^{C\sqrt{n}}} \cdot \left( \frac{1-p}{p} \right)^{C\sqrt{n}} \\
& = \frac{\left( 1+\frac{C}{p\sqrt{n}}\right)^{C\sqrt{n}}}{\left( 1-\frac{C}{(1-p)\sqrt{n}}\right)^{C\sqrt{n}}}.
\end{align*}
The last term is bounded by a constant, as
\[
\lim_{n \to \infty} \left( 1+\frac{C}{p\sqrt{n}}\right)^{C\sqrt{n}} = e^{C^2/p},
\]
and similarly
\[
\lim_{n \to \infty} \left( 1-\frac{C}{(1-p)\sqrt{n}}\right)^{C\sqrt{n}}= e^{-C^2/(1-p)}.
\]
A similar calculation works when $C<0$.
\end{proof}

\end{document}